\documentclass{article}
\usepackage{spconf,amsmath,graphicx}
\usepackage{booktabs}
\usepackage{adjustbox}

\title{VOICE CLONING: A MULTI-SPEAKER TEXT-TO-SPEECH SYNTHESIS APPROACH BASED ON TRANSFER LEARNING}
\name{Giuseppe Ruggiero$^{\star}$ \qquad Enrico Zovato$^{\star}$ \qquad Luigi Di Caro$^{\star}$ \qquad Vincent Pollet$^{\dagger}$}
  
  \address{$^{\star}$Università degli Studi di Torino \\
      $^{\dagger}$Cerence Inc.}
\begin{document}
\ninept
\maketitle
\begin{abstract}
Deep learning models are becoming predominant in many fields of machine learning. Text-to-Speech (TTS), the process of synthesizing artificial speech from text, is no exception. To this end, a deep neural network is usually trained using a corpus of several hours of recorded speech from a single speaker. Trying to produce the voice of a speaker other than the one learned is expensive and requires large effort since it is necessary to record a new dataset and retrain the model. This is the main reason why the TTS models are usually single speaker. The proposed approach has the goal to overcome these limitations trying to obtain a system which is able to model a multi-speaker acoustic space. This allows the generation of speech audio similar to the voice of different target speakers, even if they were not observed during the training phase.
\end{abstract}
\section{Introduction}
\label{sec:introduction}
Text-to-Speech (TTS) synthesis, the process of generating natural speech from text, remains a challenging task despite decades of investigation. 
Nowadays there are several TTS systems able to get impressive results in terms of synthesis of natural voices very close to human ones. Unfortunately, many of these systems learn to synthesize text only with a single voice. The goal of this work is to build a TTS system which can generate in a data efficient manner natural speech for a wide variety of speakers, not necessarily seen during the training phase. The activity that allows the creation of this type of models is called Voice Cloning and has many applications, such as restoring the ability to communicate naturally to users who have lost their voice or customizing digital assistants such as Siri.

Over time, there has been a significant interest in end-to-end TTS models trained directly from text-audio pairs; Tacotron 2 \cite{tacotron2} used WaveNet \cite{wavenet} as a vocoder to invert spectrograms generated by sequence-to-sequence with attention \cite{attention} model architecture that encodes text and decodes spectrograms, obtaining a naturalness close to the human one. 
It only supported a single speaker. Gibiansky et al. \cite{deep_voice_2} proposed a multi-speaker variation of Tacotron able to learn a low-dimensional speaker embedding for each training speaker. Deep Voice 3 \cite{deep_voice_3} introduced a fully convolutional encoder-decoder architecture which supports thousands of speakers from LibriSpeech \cite{librispeech}. However, these systems only support synthesis of voices seen during training since they learn a fixed set of speaker embeddings. Voiceloop \cite{voice_loop} proposed a novel architecture which can generate speech from voices unseen during training but requires tens of minutes of speech and transcripts of the target speaker. In recent extensions, only a few seconds of speech per speaker can be used to generate new speech in that speaker’s voice. Nachmani et al. \cite{fitting-new-speakers} for example, extended Voiceloop to utilize a target speaker encoding network to predict speaker embedding directly from a spectrogram. This network is jointly trained with the synthesis network to ensure that embeddings predicted from utterances by the same speaker are closer than embeddings computed from different speakers. Jia et al. \cite{sv2tts} proposed a speaker encoder model similar to \cite{fitting-new-speakers}, except that they used a network independently-trained exploring transfer learning from a pre-trained speaker verification model towards the synthesis model. 

This work is similar to \cite{sv2tts} however introduces different architectures and uses a new transfer learning technique still based on a pre-trained speaker verification model but exploiting utterance embeddings rather than speaker embeddings. In addition, we use a different strategy to condition the speech synthesis with the voice of speakers not observed before and compared several neural architectures for the speaker encoder model. The paper is organized as follows: Section \ref{sec:model-architecture} describes the model architecture and its formal definition; Section \ref{sec:experiments} reports experiments and results done to evaluate the proposed solution; finally conclusions are reported in Section \ref{sec:conclusion}.

\section{Model Architecture}
\label{sec:model-architecture}
\begin{figure}[t]
    \centering
    \includegraphics[width=\linewidth]{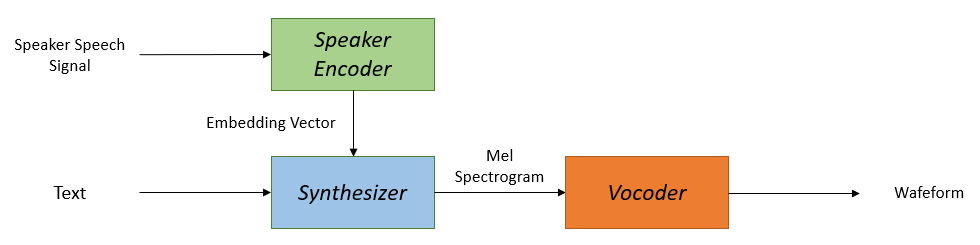}
    \caption{High level overview of the three components of the system.}
    \label{fig:model-overview}
\end{figure}

Following \cite{sv2tts}, the proposed system consists of three components: a \emph{speaker encoder}, which computes a fixed-dimensional embedding vector from a few seconds of reference speech of a target speaker; a \emph{synthesizer}, which predicts a mel spetrogram from an input text and an embedding vector; a \emph{neural vocoder}, which infers time-domain waveforms from the mel spectrograms generated by the synthesizer. 
At inference time, the speaker encoder takes as input a short reference utterance of the target speaker and generates, according to its internal learned speaker characteristics space, an embedding vector. The synthesizer takes as input a phoneme (or grapheme) sequence and generates a mel spectrogram, conditioned by the speaker encoder embedding vector. Finally the vocoder takes the output of the synthesizer and generates the speech waveform. This is illustrated in Figure \ref{fig:model-overview}.

\subsection{Problem Definition}
\label{subsec:problem-definition}
Consider a dataset of N speakers each of which has M utterances in the time-domain. Let's denote the \textit{j-th} utterance of the \textit{i-th} speaker as \textbf{u$_{ij}$} while the feature extraced from the \textit{j-th} utterance of the \textit{i-th} speaker as \textbf{x$_{ij}$} ($1 \leq i \leq N$ and $1 \leq j \leq M$). We chose as feature vector \textbf{x$_{ij}$} the mel spectrogram. 

The speaker encoder $\mathcal{E}$ has the task to produce meaningful embedding vectors that should characterize the voices of the speakers. It computes the embedding vector \textbf{e$_{ij}$} corresponding to the utterance \textbf{u$_{ij}$} as:
\begin{equation}
    \mathbf{e}_{i j}=\mathcal{E}\left(\mathbf{x}_{i j} ; \mathbf{w}_{\mathcal{E}}\right)
\label{eq:speaker_encoder_emb}
\end{equation}
where \textbf{w$_\mathcal{E}$} represents the encoder model parameters. Let's define it \emph{utterance embedding}. In addition to defining embedding at the utterance level, we can also define the \emph{speaker embedding}:
\begin{equation}
    \mathbf{c}_{i}=\frac{1}{n} \sum_{j=1}^{n} \mathbf{e}_{i j}
\label{eq:speaker_encoder_sp_emb}
\end{equation}
In \cite{sv2tts}, the synthesizer $S$ predicts \textbf{x$_{ij}$} given \textbf{c$_{ij}$} and \textbf{t$_{ij}$}, the transcript of the utterance \textbf{u$_{ij}$}:
\begin{equation}
    \hat{\mathbf{x}}_{i j}=\mathcal{S}\left(\mathbf{c}_{i}, \mathbf{t}_{i j} ; \mathbf{w}_{\mathcal{S}}\right)
    \label{eq:synth-with-speaker-emb}
\end{equation}
where \textbf{w$_\mathcal{S}$} represents the synthesizer model parameters. In our approach, we propose to use the utterance embedding rather than the speaker embedding:
\begin{equation}
    \hat{\mathbf{x}}_{i j}=\mathcal{S}\left(\mathbf{e}_{i j}, \mathbf{t}_{i j} ; \mathbf{w}_{\mathcal{S}}\right)
    \label{eq:synth-with-utt-emb}
\end{equation}
We will motivate this choice in Paragraph \ref{subsec:transfer-learning}.

Finally, the vocoder $\mathcal{V}$ generates \textbf{u$_{ij}$} given $\hat{\mathbf{\textbf{x}}}_{i j}$. So we have:
\begin{equation}
    \hat{\mathbf{u}}_{i j}=\mathcal{V}\left(\hat{\mathbf{x}}_{i j} ; \mathbf{w}_{\mathcal{V}}\right)
\end{equation}
where \textbf{w$_\mathcal{V}$} represents the vocoder model parameters. 

This system could be trained in an end-to-end mode trying to optimize the following objective function:
\begin{equation}
    \min _{\mathbf{w}_{\mathcal{E}}, \mathbf{w}_{\mathcal{S}}, \mathbf{w}_{\mathcal{V}}} L_{\mathcal{V}}\left(\mathbf{u}_{i j}, \mathcal{V}\left(\mathcal{S}\left(\mathcal{E}\left(\mathbf{x}_{i j} ; \mathbf{w}_{\mathcal{E}}\right), \mathbf{t}_{i j} ; \mathbf{w}_{\mathcal{S}}\right) ; \mathbf{w}_{\mathcal{V}}\right)\right)
\end{equation}
where \textbf{L$_{\mathcal{V}}$} is a loss function in the time-domain. However, it requires to train the three models using the same dataset, moreover, the convergence of the combined model could be hard to reach. To overcome this drawback, the synthesizer can be trained independently to directly predict the mel spectrogram \textbf{x$_{ij}$} of a target utterance \textbf{u$_{ij}$} trying to optimize the following objective function:
\begin{equation}
    \min _{\mathbf{w}_{\mathcal{S}}} L_{\mathcal{S}}\left(\mathbf{x}_{i j}, \mathcal{S}\left(\mathbf{e}_{i j}, \mathbf{t}_{i j} ; \mathbf{w}_{\mathcal{S}}\right)\right)
\end{equation}
where \textbf{L$_{\mathcal{S}}$} is a loss function in the time-frequency domain. It is necessary to have a pre-trained speaker encoder model available to compute the utterance embedding \textbf{e$_{ij}$}. 

The vocoder can be trained either directly on the mel spectrograms predicted by the synthesizer or on the groundtruth mel spectrograms:
\begin{equation}
    \min _{\mathbf{w}_{\mathcal{V}}} L_{\mathcal{V}}\left(\mathbf{u}_{i j}, \mathcal{V}\left(\mathbf{x}_{i j} ; \mathbf{w}_{\mathcal{V}}\right)\right) \text { or } \min _{\mathbf{w}_{\mathcal{V}}} L_{\mathcal{V}}\left(\mathbf{u}_{i j}, \mathcal{V}\left(\hat{\mathbf{x}}_{i j} ; \mathbf{w}_{\mathcal{V}}\right)\right)
    \label{eq:voc-two-input-types}
\end{equation}
where \textbf{L$_{\mathcal{V}}$} is a loss function in the time-domain. In the second case, a pre-trained synthesizer model is needed. 

If the definition of the objective function was quite simple for both the synthesizer and the vocoder, unfortunately this is not the case for the speaker encoder. The encoder does not have labels to be trained on because its task is only to create the space of characteristics necessary to create the embedding vectors. The Generalized End-to-End (GE2E) \cite{ge2e} loss brings a solution to this problem and it allows the training of the speaker encoder independently. Consequently, we can define the following objective function:
\begin{equation}
    \min _{\mathbf{w}_{\mathcal{E}}}
    L_{\mathcal{G}}(\mathbf{S}; w_\mathcal{E})=\sum_{j, i} L\left(\mathbf{e}_{j i}\right)
\end{equation}
where \textbf{S} represents a similarity matrix and \textbf{L$_{\mathcal{G}}$} is the GE2E loss function. 

\subsection{Speaker Encoder}
\label{subsec:speaker-encoder}
\begin{figure}[t]
    \centering
    \includegraphics[width=150pt]{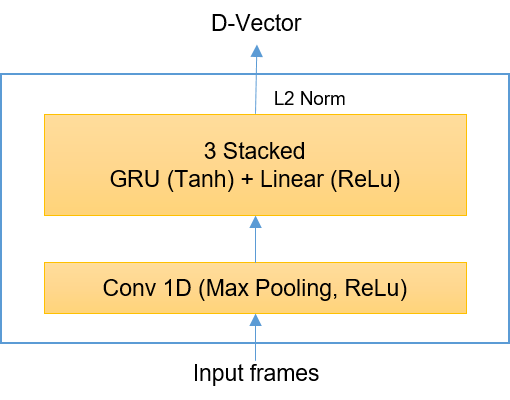}
    \caption{Speaker encoder model architecture. Input is composed of a time sequence of dimension 40. The last linear layer takes the hidden state of the last GRU layer as input.}
    \label{fig:advanced-gru-network}
\end{figure}
The speaker encoder must be able to produce an embedding vector that meaningfully represents speaker characteristics in the transformed space starting from a target speaker's utterance. Furthermore, the model should identify these characteristics using a short speech signal, regardless of its phonetic content and background noise. This can be achieved by training a neural network model on a text-independent speaker verification task that tries to optimize the GE2E loss so that embeddings of utterances from the same speaker have high cosine similarity, while those of utterances from different speakers are far apart in the embedding space. 

The network maps a sequence of mel spectrogram frames to a fixed-dimensional embedding vector, known as d-vector \cite{d-vector1, d-vector2}. Input mel spectrograms are fed to a network consisting of one Conv1D \cite{conv1D} layer of 512 units followed by a stack of 3 GRU \cite{gru} layers of 512 units, each followed by a linear projection of 256 dimension. Following \cite{sv2tts}, the final embedding dimension is 256 and it is created by L2-normalizing the output of the top layer at the final frame. This is shown in Figure \ref{fig:advanced-gru-network}. We noticed that this architecture was the best among the various tried and tested, as we will see in Section \ref{sec:experiments}. 

During the training phase, all the utterances are split into partial utterances that are 1.6 seconds long (160 frames). Also at inference time, the input utterance is split into segments of 1.6 seconds with 50\% overlap and the model processes each segment individually. Following \cite{sv2tts, ge2e}, the final utterance-wise d-vector is generated by L2 normalizing the window-wise d-vectors and taking the element-wise average.

\subsection{Synthesizer and Vocoder}
\label{subsec:synthesizer}
The synthesizer component of the system is a sequence-to-sequence model with attention \cite{tacotron2, attention} which is trained on pairs of text derived token sequences and audio derived mel spectrogram sequences. Furthermore, the network is trained in a transfer learning configuration (see Paragraph \ref{subsec:transfer-learning}), using an independently-trained speaker encoder to extract embedding vectors useful to condition the outcomes of this component. In view of reproducibility, the adopted vocoder component of the system is a Pytorch github implementation\footnote{https://github.com/fatchord/WaveRNN} of the neural vocoder WaveRNN \cite{wavernn}. This model is not directly conditioned on the output of the speaker encoder but just on the input mel spectrogram. The multi-speaker vocoder is simply trained by using data from many speakers (see Section \ref{sec:experiments}).

\subsection{Transfer Learning Modality}
\label{subsec:transfer-learning}
The conditioning of the synthesizer via speaker encoder is the fundamental part that makes the system multi-speaker: the embedding vectors computed by the speaker encoder allow the conditioning of the mel spectrograms generated by the synthesizer so that they can incorporate the new speaker voice. In \cite{sv2tts}, the embedding vectors are speaker embeddings obtained by Equation \ref{eq:speaker_encoder_sp_emb}. We used the utterance embeddings computed by Equation \ref{eq:speaker_encoder_emb}. In fact, at inference time only one utterance of the target speaker is fed to the speaker encoder which therefore produces a single utterance-level d-vector. Thus, in this case, it is not possible to create an embedding at the speaker level since the average operation cannot be applied. This implies that only utterance embeddings can be used during the inference phase. In addition, an average mechanism could cause some loss in terms of accuracy. This is due to larger variations in pitch and voice quality often occurring in utterances of the same speaker while utterances have lower intra-variation. Following \cite{sv2tts}, the embedding vectors computed by the speaker encoder are concatenated only with the synthesizer encoder output in order to condition the synthesis. However, we experimented with a new concatenation technique: first we passed the embedding through a single linear layer and then we applied the concatenation between the output of this layer and the synthesizer encoder one. The goal was to exploit the weights of the linear layer to make the embedding vector more meaningful, since the layer was trained together with the synthesizer. We noticed that this method achieved good convergence of training and was about 75\% times faster than the former vector concatenation.

\section{Experiments and Results}
\label{sec:experiments}
We used different publicly available datasets to train and evaluate the components of the system. For the speaker encoder, different neural network architectures were tested. Each of them was trained using a combination of three public sets: LibriTTS \cite{libritts} train-other and dev-other; VoxCeleb \cite{voxceleb1} dev and VoxCeleb2 \cite{voxceleb2} dev. In this way, we obtained a number of speakers equal to 8,381 and a number of utterances equal to 1,419,192, not necessarily all clean and noiseless. Furthermore, transcripts were not required. The models were trained using Adam \cite{adam} as optimizer with an initial learning rate equal to 0.001. Moreover, we experimented with different learning rate decay strategies. 

During the evaluation phase, we used a combination of the corresponding test sets of the training ones, obtaining a number of speakers equal to 191 and a number of utterances equal to 45,132. Both training and test sets have been sampled at 16 kHz and input mel spectrograms were computed from 25ms STFT analysis windows with a 10ms step and passed through a 40-channel mel-scale filterbank.

We separately trained the synthesizer and the vocoder using the same training set given by the combination of the two ``clean" sets of LibriTTS, obtaining a number of speakers equal to 1,151, a number of utterances equal to 149,736 and a total number of hours equal to 245,14 of 22.05 kHz audio. We trained the synthesizer using the L1 loss \cite{l1-loss} and Adam as optimizer. Moreover, the input texts were converted into phoneme sequences and target mel spectrogram features are computed on 50 ms signal windows, shifted by 12.5 ms and passed through an 80-channel mel-scale filterbank. The vocoder was trained using groundtruth waveforms rather than the synthesizer outputs.

\subsection{Baseline System}
\label{subsec:baseline-system}
We choose as baseline for our work the Corentin Jemine's real-time voice cloning system \cite{coren_jem_thesis}, a public re-implementation of the Google system \cite{sv2tts} available on github\footnote{https://github.com/CorentinJ/Real-Time-Voice-Cloning}. This system is composed out of three components: a recurrent speaker encoder consisting of 3 LSTM \cite{lstm} layers and a final linear layer, each of which has 256 units; a sequence-to-sequence with attention synthesizer based on \cite{tacotron2} and WaveRNN \cite{wavernn} as vocoder. 

\subsection{Speaker Encoder: Proposed System}
\label{subsec:proposed-system}
To evaluate all the speaker encoder models and choose the best one, the Speaker Verification Equal Error Rate (SV-EER) was estimated by pairing each test utterance with each enrollment speaker. The models implemented are:
\begin{itemize}
    \item \emph{rec\_conv network}: 5 Conv1D layers, 1 GRU layer and a final linear layer;
    \item \emph{rec\_conv\_2 network}: 3 Conv1D layers, 2 GRU layers each followed by a linear projection layer;
    \item \emph{gru network}: 3 GRU layers each followed by a linear projection layer;
    \item \emph{advanced\_gru network}: 1 Conv1D layer and 3 GRU layers each followed by a linear projection layer (Figure \ref{fig:advanced-gru-network});
    \item \emph{lstm network}: 1 Conv1D layer and 3 LSTM \cite{lstm} layers each followed by a linear projection layer.
\end{itemize}
All layers have 512 units except the linear ones which have 256. Moreover, dropout rate of 0.2 was used after all the layers except before the first and after the last. All the models were trained using a batch size of 64 speakers and 10 utterances for each speaker. The results obtained are shown in Table \ref{tab:summary-speaker-encoder-models-table}. 
\begin{table}[ht]
    \caption{Speaker Verification Equal Error Rates.}
    \centering
    \begin{adjustbox}{max width=\linewidth}
        \begin{tabular}{ccccc}
            \toprule
            \textbf{ Name } & {\textbf{ Step Time }} &  {\textbf{ Train Loss }} & {\textbf{ SV-EER }} & {\textbf{ LR Decay }} \\
            \midrule
            \text { rec\_conv } & \textbf{0.33s} & {0.36} & {0.073} & {Reduce on Plateau} \\
            \text { rec\_conv\_2 } & \textbf{0.45s} & {0.49} & {0.075} & {Reduce on Plateau} \\
            \text { gru } & {1,45s} & {0.33} & {0.054} & {Every 100,000 step} \\
            \text { advanced\_gru } & {0.86s} & \textbf{0.14} & \textbf{0.040} & {Exponential} \\
            \text { lstm } & {1.08s} & {0.17} & {0.052} & {Exponential} \\
            \bottomrule
        \end{tabular}
    \end{adjustbox}
    \label{tab:summary-speaker-encoder-models-table}
\end{table}

\begin{figure}[t]
    \centering
    \includegraphics[width=130pt]{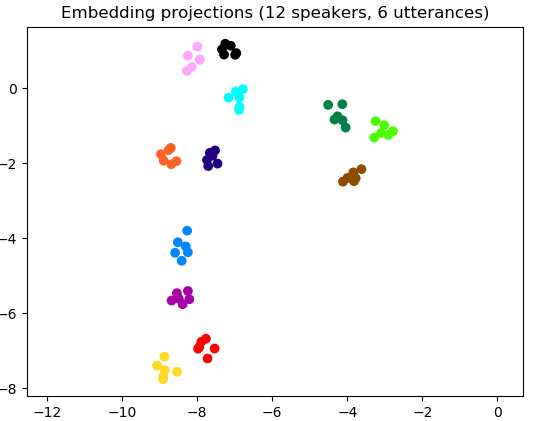}
    \caption{Advanced Gru Network test utterance embeddings projection.}
    \label{fig:adv-gru-net-test-proj1}
\end{figure}

\begin{figure}[th]
    \centering
    \includegraphics[width=130pt]{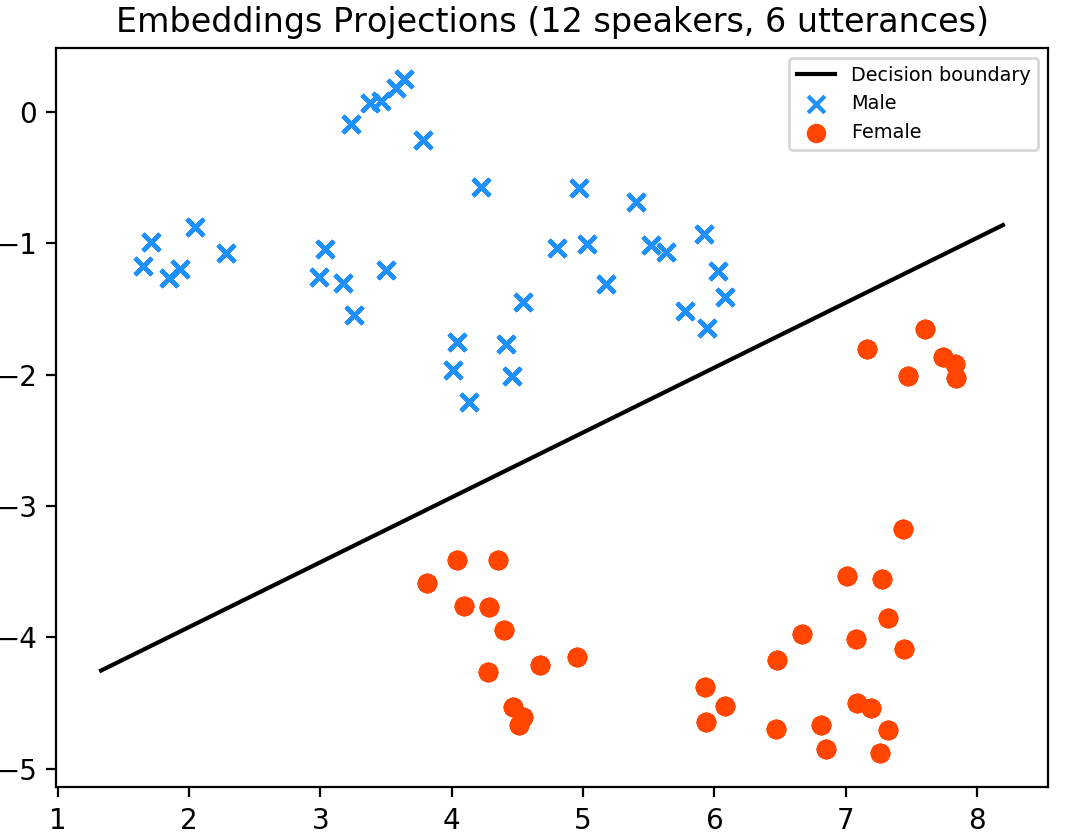}
    \caption{Advanced Gru Network six utterances for six male vs six utterances for six female taken from the test set.}
    \label{fig:adv-gru-net-test-proj2}
\end{figure}

We designed the advanced gru network trying to combine the advantages of convolutional and gru networks. In fact, looking at the table, this architecture was much faster than the gru network during training, and obtained the best SV-EER on the test set. Figure \ref{fig:adv-gru-net-test-proj1} illustrates the projection in a two-dimensional space of the utterance embeddings computed by the advanced gru network on the basis of 6 utterances extracted from 12 speakers of the test set. In Figure \ref{fig:adv-gru-net-test-proj2}, the 12 speakers are 6 men and 6 women. The projections were made using UMAP \cite{umap}. Both the figures show that the model has created a space of internal features that is robust regarding the speakers, creating well-formed clusters of speakers based on their utterances and nicely separating male speakers from female ones. 

The SV-EER obtained on the test set from the speaker encoder model of the proposed system is 0.040 vs the baseline one which is 0.049.

\subsection{Similarity Evaluation}
\label{subsec:similarity-Evaluation}
\begin{figure}[th]
    \centering
    \includegraphics[width=130pt]{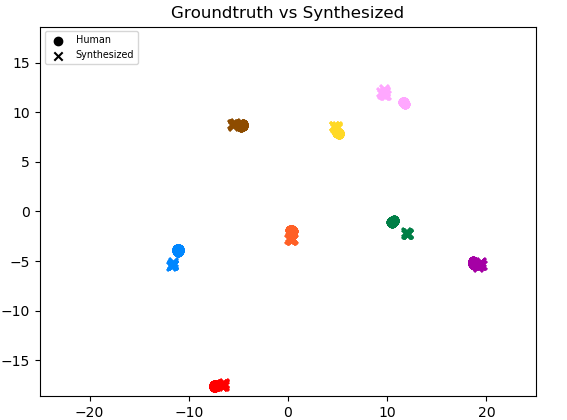}
    \caption{Groundtruth  utterance  embeddings  vs  the  corresponding  generated  ones  of  the 8 speakers chosen for testing.}
    \label{fig:voc-groundtruth-vs-ynthesized}
\end{figure}

To assess how similar the waveforms generated by the system were from the original ones, we transformed the audio signals produced into utterance embeddings (using the speaker encoder advanced gru network) and then projected them in a two-dimensional space together with the utterance embeddings computed on the basis of the groundtruth audio. As test speakers, we randomly choose eight target speakers: four speakers (two male and two female) were extracted from the test-set-clean of LibriTTS \cite{libritts}, three (two male and one female) from VCTK \cite{vctk} and finally a female proprietary voice. For each speaker we randomly extracted 10 utterances and compared them with the utterances generated by the system calculating the cosine similarity. The speakers averaged values of cosine similarity between the generated and groundtruth utterance embeddings range from 0.56 to 0.76. Figure \ref{fig:voc-groundtruth-vs-ynthesized} shows that synthesized utterances tend to lie close to real speech from the same speaker in the embedding space.

\subsection{Subjective Evaluation}
\label{subsec:mos}
Finally, we evaluated how the generated utterances were, subjectively speaking, similar in terms of speech timbre to the original ones. To do this, we gathered Mean Similarity Scores (MSS) based on a 5 points mean opinion score scale, where 1 stands for ``very different" and 5 for ``very similar". Ten utterances of the proprietary female voice were cloned using both the proposed and the baseline system and then 12 subjects, most of them TTS experts, were asked to listen to the 20 samples, randomly mixed, and rate them. Participants were also provided with an original utterance as reference. The question asked was: ``How do you rate the similarity of these samples with respect to the reference audio? Try to focus on vocal timbre and not on content, intonation or acoustic quality of the audio". The results obtained are shown in Table \ref{tab:mos-test}. Although not conclusive, this experiment highlights a subjective evidence of the goodness of the proposed approach, despite the significant variance of both systems: this is largely due to the low number of test participants.
\begin{table}[h]
    \caption{MSS of the baseline and the proposed systems.}
    \centering
    \begin{adjustbox}{max width=\linewidth}
        \begin{tabular}{cc}
            \toprule
            \textbf{ System } & \textbf{MSS} \\
            \midrule
            \text { baseline } & {$2.59 \pm 1.03$} \\
            \text { proposed } & {$3.17 \pm 0.97$} \\
            \bottomrule
        \end{tabular}
    \end{adjustbox}
    \label{tab:mos-test}
\end{table}

\section{Conclusions}
\label{sec:conclusion}
In this work, our goal was to build a Voice Cloning system which could generate natural speech for a variety of target speakers in a data efficient manner. Our system combines an independently trained speaker encoder network with a sequence-to-sequence with attention architecture and a neural vocoder model. Using a transfer learning technique from a speaker-discriminative encoder model based on utterance embeddings rather than speaker embeddings, the synthesizer and the vocoder are able to generate good quality speech also for speakers not observed before. Despite the experiments showed a reasonable similarity with real speech and improvements over the baseline, the proposed system does not fully reach human-level naturalness in contrast to the single speaker results from \cite{tacotron2}. Additionally, the system is not able to reproduce the speaker prosody of the target audio. These are consequences of the additional difficulty of generating speech for a variety of speakers given significantly less data per speaker unlike when training a model on a single speaker. 

\section{Acknowledgements}
\label{sec:acknowledgements}
The authors thank Roberto Esposito, Corentin Jemine, Quan Wang, Ignacio Lopez Moreno, Skjalg Lepsøy, Alessandro Garbo and Jürgen Van de Walle for their helpful discussions and feedback.

\bibliographystyle{IEEEbib}
\bibliography{strings,refs}

\begin{thebibliography}{10}

\bibitem{tacotron2}
J.~{Shen}, R.~{Pang}, R.~J. {Weiss}, M.~{Schuster}, N.~{Jaitly}, Z.~{Yang},
  Z.~{Chen}, Y.~{Zhang}, Y.~{Wang}, R.~{Skerrv-Ryan}, R.~A. {Saurous},
  Y.~{Agiomvrgiannakis}, and Y.~{Wu},
\newblock ``Natural tts synthesis by conditioning wavenet on mel spectrogram
  predictions,''
\newblock in {\em 2018 IEEE International Conference on Acoustics, Speech and
  Signal Processing (ICASSP)}, 2018, pp. 4779--4783.

\bibitem{wavenet}
A{\"a}ron van~den Oord, Sander Dieleman, Heiga Zen, Karen Simonyan, Oriol
  Vinyals, Alex Graves, Nal Kalchbrenner, Andrew~W. Senior, and Koray
  Kavukcuoglu,
\newblock ``Wavenet: A generative model for raw audio,''
\newblock {\em CoRR}, vol. abs/1609.03499, 2016.

\bibitem{attention}
Dzmitry Bahdanau, Kyunghyun Cho, and Yoshua Bengio,
\newblock ``Neural machine translation by jointly learning to align and
  translate,''
\newblock {\em CoRR}, vol. abs/1409.0473, 2015.

\bibitem{deep_voice_2}
Andrew Gibiansky, Sercan Arik, Gregory Diamos, John Miller, Kainan Peng, Wei
  Ping, Jonathan Raiman, and Yanqi Zhou,
\newblock ``Deep voice 2: Multi-speaker neural text-to-speech,''
\newblock in {\em Advances in Neural Information Processing Systems 30},
  I.~Guyon, U.~V. Luxburg, S.~Bengio, H.~Wallach, R.~Fergus, S.~Vishwanathan,
  and R.~Garnett, Eds., pp. 2962--2970. Curran Associates, Inc., 2017.

\bibitem{deep_voice_3}
Wei Ping, Kainan Peng, Andrew Gibiansky, Sercan~O. Arik, Ajay Kannan, Sharan
  Narang, Jonathan Raiman, and John Miller,
\newblock ``Deep voice 3: 2000-speaker neural text-to-speech,''
\newblock in {\em International Conference on Learning Representations}, 2018.

\bibitem{librispeech}
V.~{Panayotov}, G.~{Chen}, D.~{Povey}, and S.~{Khudanpur},
\newblock ``Librispeech: An asr corpus based on public domain audio books,''
\newblock in {\em 2015 IEEE International Conference on Acoustics, Speech and
  Signal Processing (ICASSP)}, 2015, pp. 5206--5210.

\bibitem{voice_loop}
Yaniv Taigman, Lior Wolf, Adam Polyak, and Eliya Nachmani,
\newblock ``Voiceloop: Voice fitting and synthesis via a phonological loop,''
\newblock in {\em International Conference on Learning Representations}, 2018.

\bibitem{fitting-new-speakers}
Eliya Nachmani, Adam Polyak, Yaniv Taigman, and Lior Wolf,
\newblock ``Fitting new speakers based on a short untranscribed sample,''
\newblock {\em CoRR}, vol. abs/1802.06984, 2018.

\bibitem{sv2tts}
Ye~Jia, Yu~Zhang, Ron~J. Weiss, Quan Wang, Jonathan Shen, Fei Ren, Zhifeng
  Chen, Patrick Nguyen, Ruoming Pang, Ignacio Lopez{-}Moreno, and Yonghui Wu,
\newblock ``Transfer learning from speaker verification to multispeaker
  text-to-speech synthesis,''
\newblock {\em CoRR}, vol. abs/1806.04558, 2018.

\bibitem{ge2e}
Lipeng Wan, Qi~shan Wang, Alan Papir, and Ignacio Lopez-Moreno,
\newblock ``Generalized end-to-end loss for speaker verification,''
\newblock {\em 2018 IEEE International Conference on Acoustics, Speech and
  Signal Processing (ICASSP)}, pp. 4879--4883, 2018.

\bibitem{d-vector1}
Georg Heigold, Ignacio Moreno, Samy Bengio, and Noam Shazeer,
\newblock ``End-to-end text-dependent speaker verification,''
\newblock {\em 2016 IEEE International Conference on Acoustics, Speech and
  Signal Processing (ICASSP)}, pp. 5115--5119, 2016.

\bibitem{d-vector2}
Ehsan Variani, Xin Lei, Erik McDermott, Ignacio~Lopez Moreno, and Javier
  Gonzalez-Dominguez,
\newblock ``Deep neural networks for small footprint text-dependent speaker
  verification,''
\newblock in {\em Proc. ICASSP}, 2014.

\bibitem{conv1D}
Serkan Kiranyaz, Onur Avci, Osama Abdeljaber, Turker Ince, Moncef Gabbouj, and
  Daniel~J. Inman,
\newblock ``1d convolutional neural networks and applications: A survey,''
\newblock {\em ArXiv}, vol. abs/1905.03554, 2019.

\bibitem{gru}
Junyoung Chung, Caglar Gulcehre, Kyunghyun Cho, and Yoshua Bengio,
\newblock ``Empirical evaluation of gated recurrent neural networks on sequence
  modeling,''
\newblock in {\em NIPS 2014 Workshop on Deep Learning, December 2014}, 2014.

\bibitem{wavernn}
Nal Kalchbrenner, Erich Elsen, Karen Simonyan, Seb Noury, Norman Casagrande,
  Edward Lockhart, Florian Stimberg, A{\"a}ron van~den Oord, Sander Dieleman,
  and Koray Kavukcuoglu,
\newblock ``Efficient neural audio synthesis,''
\newblock in {\em ICML}, 2018.

\bibitem{libritts}
Heiga Zen, Viet Dang, Rob Clark, Yu~Zhang, Ron~J. Weiss, Ye~Jia, Zhifeng Chen,
  and Yonghui Wu,
\newblock ``Libritts: A corpus derived from librispeech for text-to-speech,''
\newblock in {\em INTERSPEECH}, 2019.

\bibitem{voxceleb1}
Arsha Nagrani, Joon~Son Chung, and Andrew Zisserman,
\newblock ``Voxceleb: A large-scale speaker identification dataset,''
\newblock in {\em INTERSPEECH}, 2017.

\bibitem{voxceleb2}
Joon~Son Chung, Arsha Nagrani, and Andrew Zisserman,
\newblock ``Voxceleb2: Deep speaker recognition,''
\newblock in {\em INTERSPEECH}, 2018.

\bibitem{adam}
Diederik Kingma and Jimmy Ba,
\newblock ``Adam: A method for stochastic optimization,''
\newblock {\em International Conference on Learning Representations}, 12 2014.

\bibitem{l1-loss}
Katarzyna Janocha and Wojciech Czarnecki,
\newblock ``On loss functions for deep neural networks in classification,''
\newblock {\em ArXiv}, vol. abs/1702.05659, 2017.

\bibitem{coren_jem_thesis}
Corentin Jemine,
\newblock ``Master thesis: Automatic multispeaker voice cloning,'' 2019,
\newblock Unpublished master's thesis, Université de Liège, Liège, Belgique.

\bibitem{lstm}
Klaus Greff, Rupesh~K. Srivastava, Jan Koutnik, Bas~R. Steunebrink, and Jurgen
  Schmidhuber,
\newblock ``Lstm: A search space odyssey,''
\newblock {\em IEEE Transactions on Neural Networks and Learning Systems}, vol.
  28, no. 10, pp. 2222–2232, Oct 2017.

\bibitem{umap}
Leland McInnes and John Healy,
\newblock ``Umap: Uniform manifold approximation and projection for dimension
  reduction,''
\newblock {\em ArXiv}, vol. abs/1802.03426, 2018.

\bibitem{vctk}
Christophe Veaux, Junichi Yamagishi, and Kirsten MacDonald,
\newblock ``Cstr vctk corpus: English multi-speaker corpus for cstr voice
  cloning toolkit,'' 2018.

\end{thebibliography}

\end{document}